\documentclass[traditabstract]{aa}
\usepackage{color}
\usepackage{txfonts,epsfig,graphicx,natbib,twoopt}
\bibpunct{(}{)}{;}{a}{}{,}

\begin{document}
   \title{The discovery based on GLIMPSE data of a protostar \\
   driving a bipolar outflow}

   \author{Jing-Hua Yuan\inst{1}, Jin Zeng Li\inst{1}, Ya Fang Huang\inst{1}, Chih-Hao Hsia\inst{2}
        \and Jingqi Miao\inst{3}}

   \titlerunning{A Protostar Driving a Bipolar Outflow}
    \authorrunning{J. H.~Yuan et al.}
    \institute{National Astronomical Observatories, Chinese Academy of Sciences,20A Datun Road, Chaoyang District, Beijing 100012, China\\
            \email{jhyuan@nao.cas.cn}
        \and
            Department of Physics, The University of Hong Kong, Hong Kong, China
        \and
            Centre for Astrophysics and Planetary Science, School of Physical Sciences,University of Kent, Canterbury, Kent CT2 7NR, UK}

    \abstract{We report the discovery based on GLIMPSE data of a proto-stellar system driving a bipolar outflow . The bipolar outflow closely resembles the shape of an hourglass in the infrared. The total luminosity of $L_{total}=5507$ L$_\odot$, derived from IRAS fluxes, indicates the ongoing formation of a massive star in this region. The spectral energy distribution (SED) of the driving source is fitted with an online SED fitting tool, which results in a spectral index of about 1.2. This, along with the presence of a bipolar outflow, suggests the detection of a Class I protostar. The driving source indicates prominent infrared excesses in color-color diagrams based on archived 2MASS and GLIMPSE data, which is in line with an early evolutionary stage of the system.}

    \maketitle

\section{Introduction}

    The mass of a young stellar object (YSO) is the key factor that determines its evolutionary path after its formation. Young stellar objects with masses below and above 8 M$_{\odot}$ experience substantially different processes in their formation \citep{zin07,eva11}. For low-mass star formation, a widely accepted scenario has been proposed by \citet{shu87}, in which four distinct evolutionary stages were defined. On the basis of the star formation scenario proposed by \citet{shu87}, \citet{lad87} developed a widely used classification scheme for YSOs primarily based on their spectral energy distributions (SEDs) and the spectral index at the infrared. With an evolutionary sequence from early to late type, YSOs were classified into Class I to III \citep{lad87}. Later on, an earlier evolutionary phase called Class 0 was added to this scheme by \citet{and94}.

    However, the scenario of massive star formation remains unclear \citep{zin07}. Two competing schemes are well known in the literature. As their low-mass counterparts, massive stars were proposed to form via accretion of ambient gas, though the process is much more violent \citep{gar99}. On the other hand, massive stars were alternatively proposed to form by merging of low-mass stars at the center of dense stellar clusters \citep{bon98}. As observational evidence accumulates, the accretion scenario turns out to be the more likely one.

    Outflows are commonly associated with both low-mass and massive star formation \citep{lad85,bac96,wu04,arc07}. Massive outflows, as the feedback of star formation, may create strong impact on the ambient conditions such as chemical abundance and density, the final mass of the central star in formation and star-forming activities in both of the close vicinity and far reaches.

    In this paper, we report on a newly discovered Class I-type protostar driving an hourglass-shaped outflow. Little is known in the literature about the likely exciting source GLIMPSE G012.4013-00.4687, located at (R.A.=$18^h 14^m 24^s.56$, Dec.=$-18^\circ24^\prime42^{\prime \prime}.69$ J2000). The associated 2MASS point source given in the highly reliable GLIMPSE catalog is 2MASS J18142457-1824421. And the 2MASS photometric data reveal high infrared excess. An IRAS source (IRAS 18114-1825) \citep{cas89} is located $9.3^{\prime\prime}$ to its northwest. \citet{cod95} and \citet{mac98} tried to search for water and 6.7 GHz methanol masers toward IRAS 18114-1825 but achieved negative results. The nature of IRAS 18114-1825 thus remains unclear.

    This paper is arranged as follows: we present a description of the archival data and observations in Section \ref{s2}, while the detection of the bipolar outflow system and the subsequent analysis are described in Section \ref{s3}. In Section \ref{s4} we exclude the possibility of an evolved object. The evolutionary stage of this source is investigated in Section \ref{s5} and the results are summarized in Section \ref{s6}.

\section{Date acquisition and analysis}\label{s2}
    \subsection{Archival data}

    GLIMPSE (the Galactic Legacy Infrared Mid-Plane Survey Extraordinaire) is a legacy project of the \textit{Spitzer Space Telescope} \citep{wer04}. The Infrared Array Camera (IRAC) \citep{faz04} was employed in this survey, which covers four infrared wavebands centered at 3.6, 4.5, 5.8, and 8 $\mu$m, respectively. Archived GLIMPSE I \citep{ben03} data were used in this work. A composite color image was compiled from the imaging data at 3.6 $\mu$m (blue), 4.5 $\mu$m (green) and 8 $\mu$m (red). Photometric data from the highly reliable GLIMPSE I catalog were used to fit the SED of the exciting source. We only considered sources with data available in all of the four IRAC bands.

    Archived data from the 2MASS Point Source Catalog (PSC) were also used in our work. To guarantee the reliability of the data, we adopted the source selection criteria from \citet{li05}. These include that a) each source extracted from the 2MASS PSC must have a certain detection in all $J,H$ $\&$ $Ks$ wavebands; b) only sources with a $Ks$-band signal-to-noise ratio above 15 are selected. The retrieved 2MASS PSC data were used in the SED fitting and to obtain the near-infrared color-color diagram.

    The Bolocam Galactic Plane Survey \citep[BGPS, ][]{bal10,ros10} is a 1.1 mm continuum survey of the Galactic Plane performed with the Bolocam on the Caltech Submillimeter Observatory, which is operated by Caltech under a contract from the NSF. A cutout image from archived Bolocam 1.1 mm data was employed in this investigation.

    Photometric data from MSX, DENIS and IRAS were also included in the construction and the fitting of the SED of the central exciting source.

    \subsection{Optical spectroscopy}
    Low-resolution spectroscopy of the central source of the bipolar outflow was performed on the nights of 11 and 13 September 2010 with the 2.16 m telescope of the National Astronomical Observatories, Chinese Academy of Sciences. An OMR (Optomechanics Research Inc.) spectrograph and a PI $1340\times400$ CCD were used. Here, a 200 {\AA}mm$^{-1}$ grating and a 2$^{\prime\prime}$ slit, across the source along the declination direction (P.A.=0$^\circ$), resulted in a two-pixel resolution of 9.6 {\AA}.  A standard star, HR 7596, was observed as the flux calibrator.  Because the source was faint in the optical, we took three exposures, one on 11 September and the other two on 13 September, and then combined them to get a better signal-to-noise ratio. The exposure time lasted an hour for each exposure.

\section{Results}\label{s3}
    \begin{figure*}
    \begin{center}
    \includegraphics[width=0.8\textwidth]{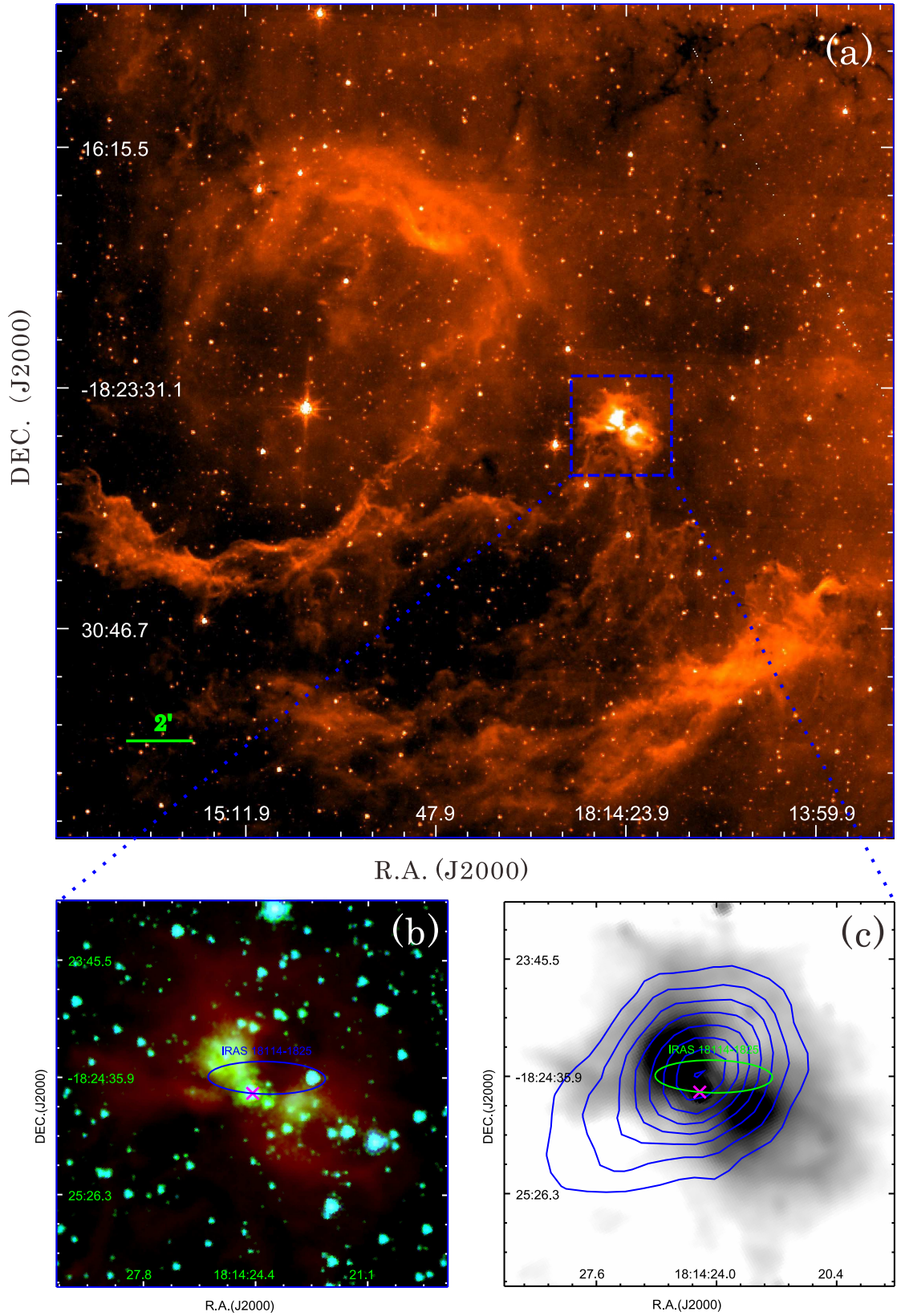}

    \end{center}
    \caption{(\emph{a}):\emph{Spitzer}/IRAC 8 $\mu$m mosaic of the bubble N6 in \citet{deh10}. The blue dashed box indicates the the outflow structure. (\emph{b}): The composite color image of the IRAC mosaics of the bipolar system, where the images at 8, 4.5, and 3.6 $\mu$m  are mapped in red, green and blue respectively. (\emph{c}): Gray-scaled color-inverted \textit{Spitzer}/MIPS 24 $\mu$m image is overlapped with the contours of Bolocam 1.1 mm continuum emission. The intensity of the contours varies from 0.1 Jy/Beam to 0.8 Jy/Beam with 10 levels. The error ellipse of the IRAS point source (IRAS 18114-1825) is over-plotted on both panels. The magenta cross marks the central source of the system.}
    \label{fig1}
    \end{figure*}

    Fig. \ref{fig1}(\emph{a}) presents the \emph{Spitzer}/IRAC 8 $\mu$m mosaic of a $25^\prime\times25^\prime$ region, in which the outflow structure is marked with a blue box. It is located at the junction of two large filaments. These filaments are recently reported as potential regions of active star formation in theoretical and observational studies \citep{dal11,and10,mol10,hil11}. In Fig. \ref{fig1}(\emph{a}), the eastern extended cavity with a bright central star and the southern filamentary structure are parts of a complex bubble, N6 as presented by \citet{deh10}. The nature of N6 is still unrevealed. The bipolar source of interest here is situated between the two components of N6 and looks to be linked to both with dust showing week PAH emissions. This could be due to the projection effect. Additional observations with molecular lines at longer wavelength are necessary to make this clear.

    As shown in the composite image (Fig. \ref{fig1} \emph{b}), compiled with IRAC 8 $\mu$m (red), 4.5 $\mu$m (green), and 3.6 $\mu$m (blue) mosaics, the GLIMPSE data reveal an hourglass-shaped structure with a NE-SW orientation (P.A.=$52^{\circ}$). The prominent extended emission of 4.5 $\mu$m, containing both H$_2$(v=0-0, s(9,10,11)) lines and the CO(v=1-0) band head, suggests the detection of a massive young stellar object with ongoing outflow activity \citep{cyg08}. The probable exciting source, GLIMPSE G012.4013-00.4687, is marked with a magenta cross, which is located just at the edge of the error ellipse of IRAS 18114-1825. This suggests an association of the exciting source of the bipolar structure with IRAS 18114-1825.

    The bipolar structure is also detected at \emph{Spitzer}/MIPS 24 $\mu$m. The Bolocam 1.1 mm dust continuum reveals an elongated structure perpendicular to the bipolar system, which is interpreted here as a massive envelope and will be discussed in detail in  \S\ref{s3.2}.

    \subsection{Distance}
    The interstellar extinction to the infrared point source is calculated to be $A_V=2.384$ mag, according to its 2MASS colors (e.g., $J-H$ and $H-K_s$). A distance of  2.15 Kpc is derived based on an extinction model toward the Galactic coordinate of the bipolar system developed by \citet{amo05}.

    With the extinction law ($A_{K_s}/A_V=0.112$) proposed by \citet{rie85}, we obtained the extinction in the $K_s$ band, $A_{K_s}=0.267$ mag. A distance of 2.67 Kpc was estimated based on the model of \citet{mar06}. The averaged value (d=2.41$\pm$0.26 Kpc) of the distance estimated with the two different models was then adopted.

    \subsection{Physical parameters}\label{s3.2}
    The total flux density of the infrared point source can be derived from the IRAS data with the following equation \citep{cas86}:
    \begin{equation}
    F(10^{-13} \mathrm{W m^{-2}})=\frac{1.75}{\frac{F_{12}}{0.79}+\frac{F_{25}}{2}+\frac{F_{60}}{3.9}+\frac{F_{100}}{9.9}},
    \end{equation}
   where $F_{12}$, $F_{25}$, $F_{60}$, and $F_{100}$ are the flux densities in Jy at 12 $\mu$m, 25 $\mu$m, 60 $\mu$m, and 100 $\mu$m respectively. With the estimated distance of 2.41 Kpc, we obtain a total luminosity of $L_{total}=5507\pm1200$ L$_{\odot}$. A luminosity this high suggests the ongoing formation of a massive star or a quite massive stellar system in this region.

    A dust temperature of T$_d\approx29.8$ K was estimated from the color temperature T$_c$ defined by \citet{hen90},
    \begin{equation}
    T_d \approx T_c(60/100)=\frac{96}{(3+\beta)\mathrm{ln}(\frac{100}{60})-\mathrm{ln}(\frac{F_{60}}{F_{100}})},
    \end{equation}
     where $F_{60}$ and $F_{100}$ are the flux densities in Jy at 60 $\mu$m and 100 $\mu$m respectively, and $\beta$ the emissivity index of dust particles, which is set to be 2.

    A BGPS source is detected with Bolocam at 1.1mm (see Fig. \ref{fig1} \emph{c}). With an assumption of an isothermal condition in this region, the mass of the BGPS source is obtained by adopting equation (10) of \citet{ros10},
    \begin{eqnarray}
    M = \frac{d^2S}{B_\nu(T)\kappa_\nu} = 13.1M_{\odot}(\frac{d}{1\ \mathrm{Kpc}})^2(\frac{S_\nu}{1\ \mathrm{Jy}})[\frac{\mathrm{exp}(\frac{13.0\ \mathrm{K}}{T_d})-1}{\mathrm{exp}(\frac{13.0}{20})-1}],
    \end{eqnarray}
    where $S_\nu$ is the flux density measurement in the BGPS catalog, $B_\nu$ is the Planck function and the opacity $\kappa_\nu$ is assumed as 0.0114 $\mathrm{cm^2g^{-1}}$ \citep{eno06}. The estimated mass of the BGPS source is about $69\pm15$ M$_{\odot}$, which is in line with the estimated mass of the envelope ($M_{env}\approx22\ \mathrm{M_\odot}$, see \S\ref{s5.2}) by the SED fitting. This substantiates the detection of a massive envelope in association with the outflow system.

    \begin{figure}
    \begin{center}
    \includegraphics[width=0.5\textwidth]{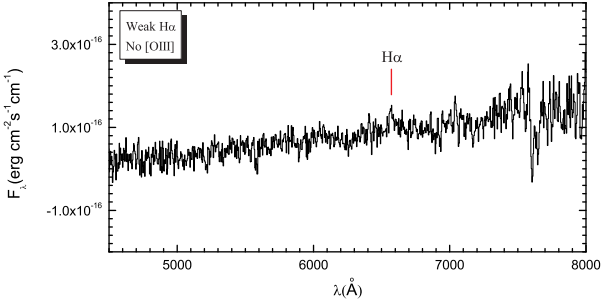}
    \end{center}
    \caption{Low-resolution spectrum of the central source of the bipolar system.}
    \label{fig2}
    \end{figure}

\section{Young stellar object vs. evolved stellar object}\label{s4}

    We cannot exclude, however, its possible origin in a young planetary nebula (PN) or a proto-planetary nebula (PPN) because these also show similar bipolar morphologies in the infrared \citep{bal87,kwo93}. Recombination and forbidden line emission from hydrogen and metals, especially H$_\alpha$ and [O\textsc{iii}], are significant spectral features of a PN in the optical \citep{ost64,mil74}. However, very weak H$_\alpha$ and no [O\textsc{iii}] emission were detected in the optical spectrum (see Fig. \ref{fig2}), which indicates that the bipolar system is not a PN. On the other hand, a mass estimate of about 9.2 M$_\odot$ of the central exciting source (see \S\ref{s5.2}) is much higher than that expected for a PN or a PPN, which is usually $\sim$1 M$_\odot$ \citep{kal85}.

    The major diameter of the bipolar structure is about 110$^{\prime\prime}$, corresponding to a physical scale of 1.29 pc at the distance of 2.41 kpc. Given an expansion velocity of 30 km s$^{-1}$ \citep[the typical speed of expansion of a PPN ranges from a few to 30 km s$^{-1}$,][]{kwo93}, a dynamic age of $2.096\times10^4$ yr is calculated, which is inconsistent with the age of a PPN that ranges from a few hundred to several thousand years \citep{kwo93}. Another supporting evidence is the mass of the envelope derived from the Bolocam 1.1 mm data (see \S\ref{s3.2}), which is far beyond that possible for material ejected from an evolved star.

    Compared with that of their pre-main-sequence counterparts (Class II and III), the optical spectral properties of protostars  (Class I) have seldom been studied \citep{whi07}. \citet{whi04} carried out spectroscopic investigation into 15 Class I YSOs in the optical domain, in which strong emission lines (e.g., H$_\alpha$) are detected. And P-cygni profiles of broad H$_\alpha$ emission of some sources reveals the features of outflows or jets initiated from the central protostars or disks. In some of  \citet{whi04}'s sample,  forbidden lines were also detected, e.g., [S \textsc{ii}] $\lambda$6731, whose equivalent width is always used for the estimation of mass outflow rates. In our observations of the likely exciting source, only weak H$_\alpha$ is traced, and no forbidden emissions (e.g.,[S \textsc{ii}] $\lambda$6731) are detected (see Fig. \ref{fig2}). The deep embedded and young nature of the massive protostar could be responsible for the weakness of H$_\alpha$ and the absence of [S \textsc{ii}] $\lambda$6731. Observations with a higher sensitivity and a slit along the outflow axis (P.A.=$52^{\circ}$) are needed for a closer investigation into its optical spectrum.

\section{Evolutionary status}\label{s5}
    \begin{figure}
    \begin{center}
    \includegraphics[width=0.45\textwidth]{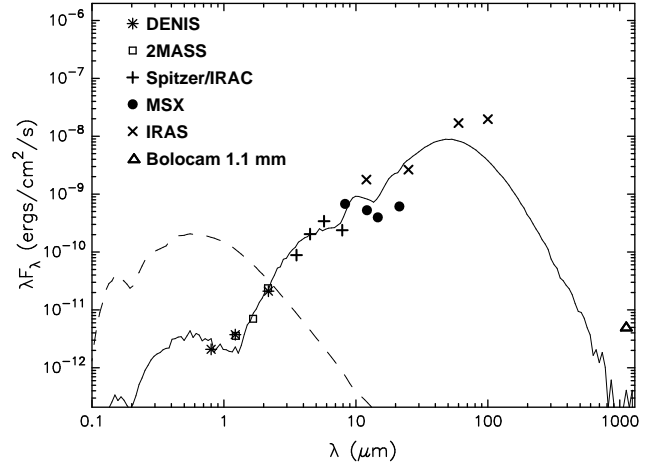}
    \end{center}
    \caption{Fitted SED of the protostar. The solid line shows the best-fitted SED model. Variant symbols represent near- to far-infrared photometric data of the protostar, while the dashed line is for the SED of the forthcoming stellar photosphere in the best-fitted model \citep{rob07}.}
    \label{fig3}
    \end{figure}

    \begin{figure*}
        \begin{center}
        \includegraphics[width=0.33\textwidth]{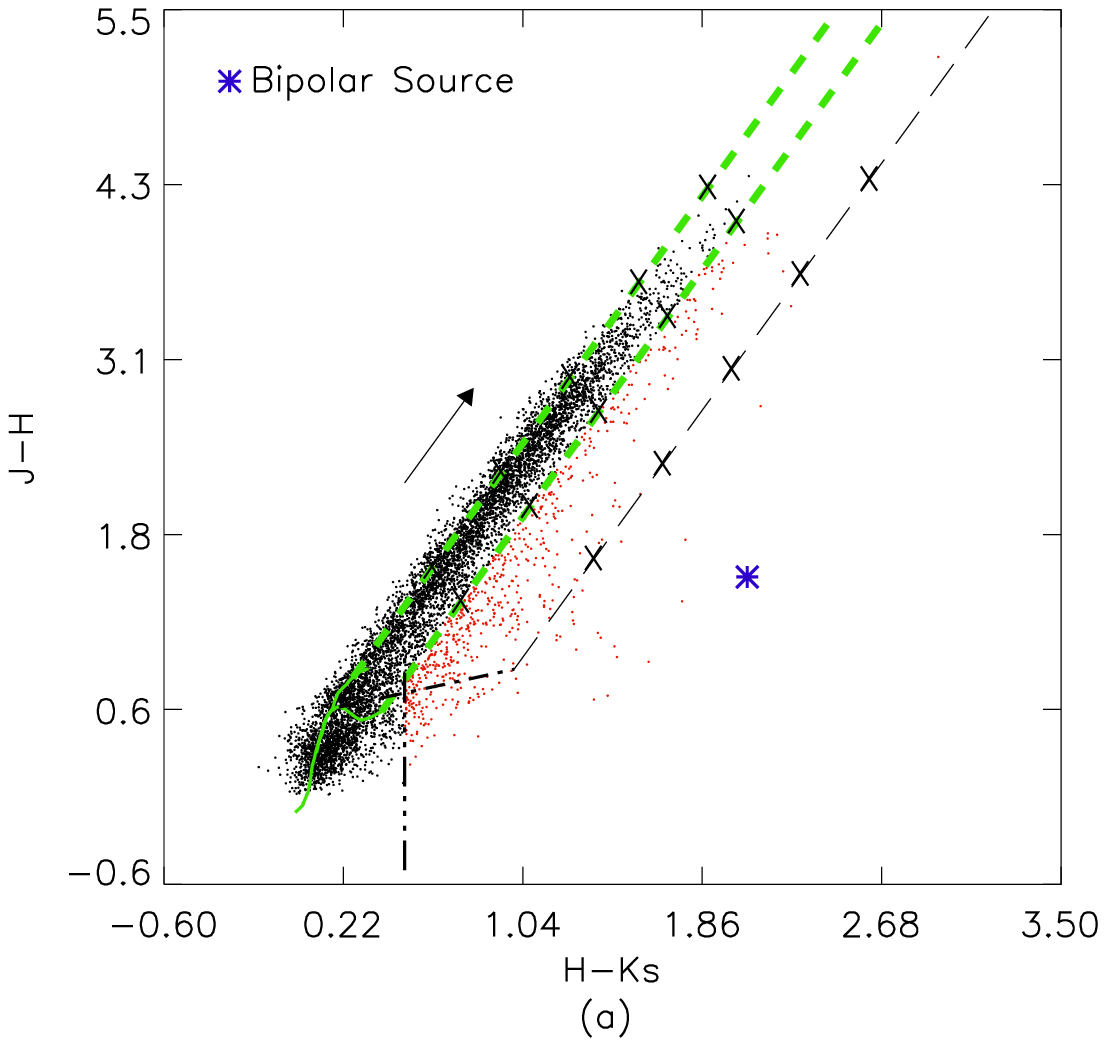}
        \includegraphics[width=0.33\textwidth]{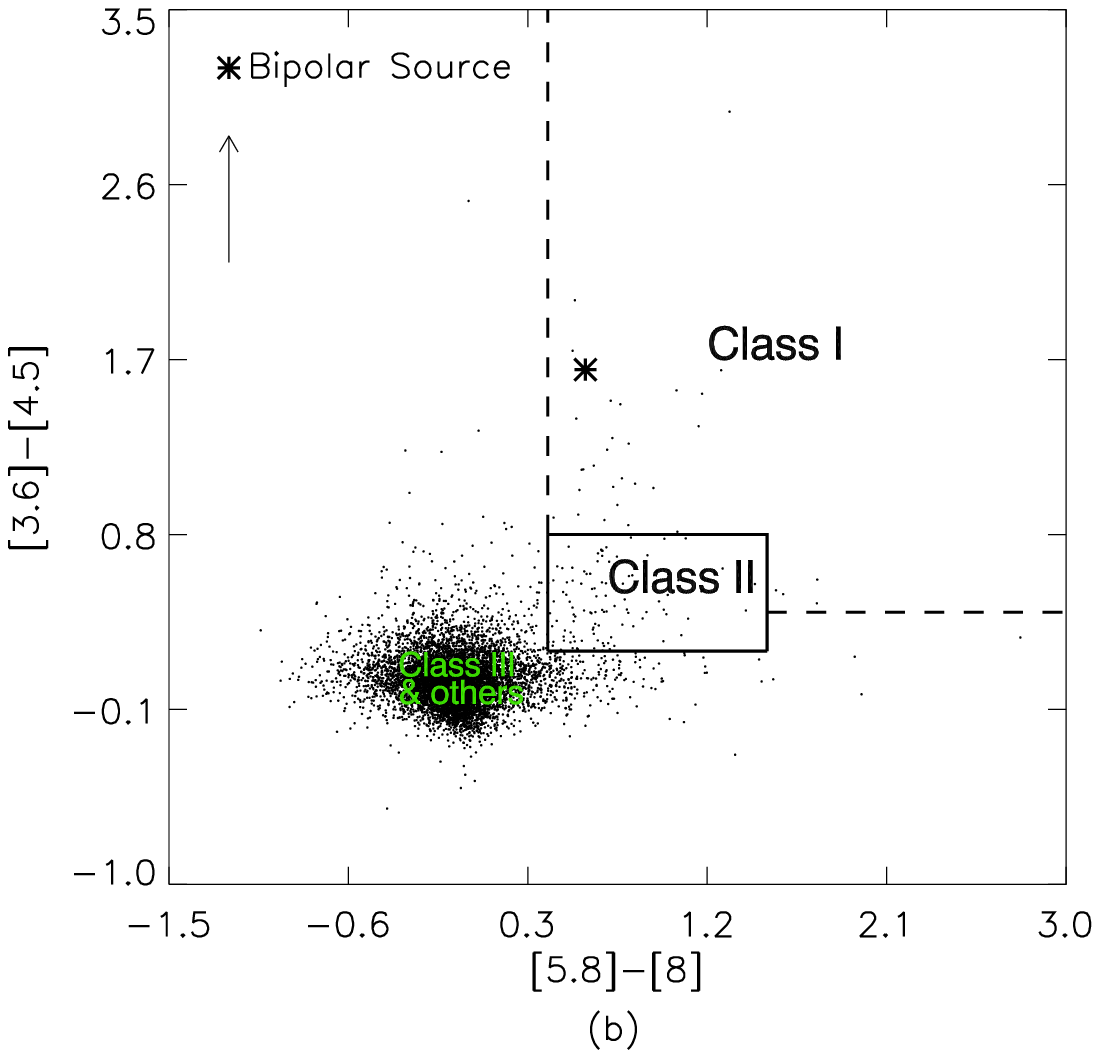}
        \includegraphics[width=0.33\textwidth]{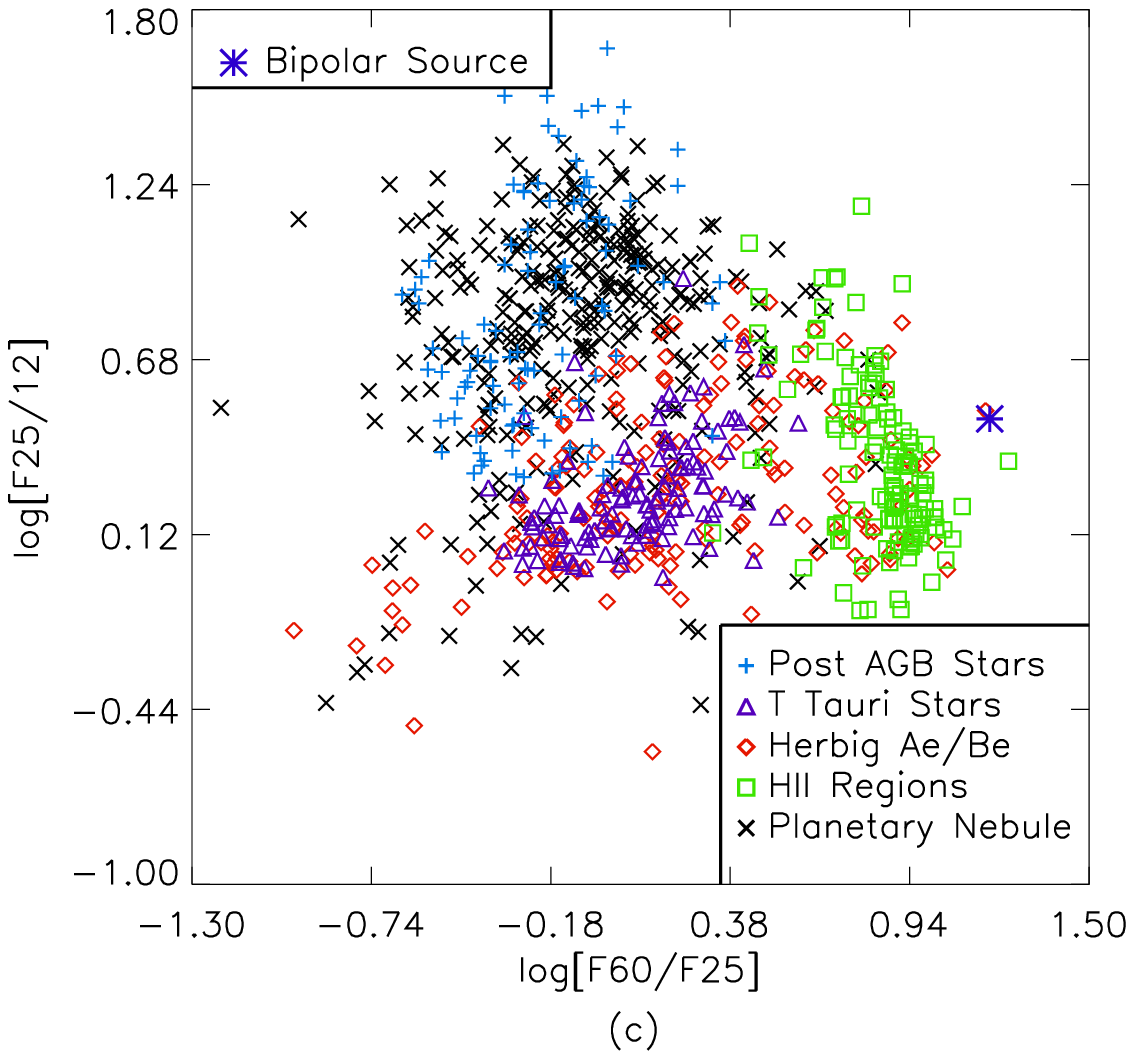}
        \end{center}
    \caption{Near- to far-infrared color-color diagrams. (\emph{a}): ($J-H$) vs. ($H-K_s$) diagram. Green solid lines are the loci of the main-sequence dwarfs and giant stars \citep{bes88}. The arrow shows a reddening vector of $A_V =5$ mag \citep{rie85}. The dot-dashed line indicates the locus of dereddened T Tauri stars \citep{mey97}. The dashed lines, which are drawn parallel to the reddening vector, define the reddening band for normal field stars and T Tauri stars. Crosses are overplotted with an interval corresponding to $A_V =5$ mag. (\emph{b}): [3.6]-[4.5] vs. [5.8]-[8] diagram. The classification scheme is based \citet{meg04} and \citet{all04,all07}. The arrow indicates an extinction of $A_{Ks}=5$ mag according to \citet{ind05}. (\emph{c}): log[F60/F25] vs. log[F25/F12] diagram. The sample of H\textsc{ii} regions, T Tauri stars, Herbig Ae/Be stars, post AGB stars, and planetary nebulae are adopted from \citet{pal91}, \citet{zha06a}, \citet{zha06b}, \citet{sua06}, and \citet{taj98}, respectively. The protostar is presented as an asterisk in all three panels. }
    \label{fig4}
    \end{figure*}

    \subsection{Implications of the morphology}

   Previous studies indicate the opening angle of the outflows associated with massive YSOs tends to increase with evolutionary age \citep{arc06,sea08}. For example, outflows associated with Class 0 sources used to be highly collimated, while those in association with Class I protostars are found to have poor collimation \citep{arc07}. The opening angle of the outflow reported here  is estimated to be about $101^\circ$, and the collimation factor (the ratio of major to minor radii) is about 2. This poor collimation, together with the large opening angle, indicates that the central YSO cannot be in its very early stages of evolution.

   The investigation of a jet-like bipolar outflow was reported by \citet{var11}. Near-infrared emissions, at \emph{JHK} and IRAC 4.5$\mu$m wavebands, reveal its well-collimated feature, which was interpreted as evidence for its accretion-driven nature \citep{var11}. In contrast, the outflow discussed here is not well-collimated, and its engine is still unclear. One possible origin of poorly collimated outflows like the one presented here is the entrainment in the adjacent medium from high-velocity jets originated from disk winds or x-winds \citep[e.g.,][]{arc07}. However, recent numerical studies suggest different driving mechanisms for outflows with and without good collimation \citep{mac08}. \citet{mac08} found two distinct flows: a low-velocity flow with a wide opening angle, driven from the adiabatic core; and a high-velocity flow with good collimation, driven from the protostar. And this scenario is supported by other previous theoretical works \citep[e.g.,][]{tom02,ban06}. And the hourglass-shaped outflow reported here resembles that presented in \citet{mac08}. Observations with outflow-tracing molecules (e.g., CO, SiO) will give clues to its genuine driving mechanism.

    \subsection{Spectral energy distribution}\label{s5.2}
    Based on the multi-wavelength photometric data, the SED of the central YSO was calculated and fitted with an online tool\footnotemark[1] developed by \citet{rob06,rob07}. \citet{rob06} presented a grid of radiation transfer models of YSOs,  covering a wide range of possible evolutionary stages and stellar masses. The grid consists of 20,000 YSO models resulting in a total of 200,000 SEDs \citep{rob06}.

    \footnotetext[1]{http://caravan.astro.wisc.edu/protostars}

    The best-fitted SED of the central protostar is shown as a solid line in Fig. \ref{fig3}. The resulting SED resembles that of a Class I object well \citep{and94}. The fitting also indicates a central stellar mass of $\sim9.2\pm0.8$ M$_\odot$ and a total luminosity of $(4\pm1)\times10^3$ L$_\odot$, consistent with the result of 5507 L$_\odot$ estimated from the IRAS fluxes. The spectral index at infrared is calculated to be 1.2, which is consistent with that of Class I objects \citep{lad87}.

    \subsection{Color-color diagrams}

    Fig. \ref{fig4} (\emph{a}) is the ($J-H$) vs. ($H-Ks$) color-color diagram of the 2MASS point sources in a $30^{\prime}\times30^{\prime}$ region surrounding the bipolar system. All sources with excessive emission in the near-infrared are plotted in red. The exciting source of the bipolar system is presented as an asterisk, which shows prominent excessive emission in the infrared and is in line with the protostellar nature of the system.

    Fig. \ref{fig4} (\emph{b}) is the [3.6]-[4.5] vs. [5.8]-[8] map plotted with cataloged point sources from GLIMPSE I within the same region. The sample sources are shown as black dots. Sources with colors of $0.2\leq[3.6]-[4.5]\leq0.8$ and  $0.4\leq[5.8]-[8]\leq1.5$ are taken as Class II-type, while those with colors of $[3.6]-[4.5]\geq0.4$ and $[5.8]-[8]\geq0.4$ as Class-I type \citep{meg04,all04,all07}. Presented as an asterisk, the exciting source of the bipolar system is confirmed as a Class I source.

    A sample of evolved and young stellar objects are plotted on a log[F60/F25] vs. log[F25/F12] diagram (Fig. \ref{fig4} \emph{c}). The central source of the bipolar system,  presented as a blue asterisk, is located in a region dominated by young stellar objects, H\textsc{ii} regions and/or Herbig Ae/Be stars. The position of the exciting source of the bipolar nebula provides additional evidence for the protostellar nature of the system.

\section{Conclusions}\label{s6}

     We reported the discovery of an hourglass-shaped nebula at \, (R.A.=$18^h 14^m 24^s.56$,\, Dec.=$-18^\circ24^\prime42^{\prime \prime}.69$ J2000) based on GLIMPSE data. The bipolar system is oriented in the north-east and south-west direction. A distance of 2.41 kpc is estimated based on galactic extinction models. A dust temperature of $T_d=29.8$ K and a luminosity of $L_{total}=5507$ L$_{\odot}$ are derived from the IRAS fluxes, which suggests the ongoing formation of a massive star or a quite massive stellar system in this region.

    Archived Bolocam 1.1 mm data, tracing cold dust, indicate the existence of an extensive structure at the waist of the bipolar system. The mass of the BGPS source is estimated to be around 69 M$_\odot$, which is in line with the mass estimate of the envelope from the SED fitting ($M_{env}\approx22$ M$_\odot$). This substantiates the detection of a massive envelope associated with the outflow system.

    The SED fitting results in a spectral index of 1.2 in the near- to far-infrared, which, along with the high infrared excess, strongly suggests the detection of a protostellar system. A
   stellar mass of $9.2\pm0.8$ M$_\odot$ and a luminosity of $(4\pm1)\times10^3$ L$_\odot$ is obtained, which is consistent with the 5507 L$_\odot$ estimated from the IRAS fluxes.

    This study substantiates the discovery of a massive protostellar system driving a bipolar outflow. Future investigations with higher spatial resolution may provide helpful information on the the kinematics, dynamics and chemistry of the inner and denser region, and contribute to our understanding of massive star formation in its early stages.

\begin{acknowledgements}
    We thank the anonymous referee for the constructive comments that helped us improve this paper.
    This work used archived GLIMPSE and 2MASS data. Our investigation is supported by funding from the National Natural Science Foundation of China through grant NSFC 11073027 and the Department of International Cooperation of the Ministry of Science and Technology of China through grant 2010DFA02710. C.H. is supported by a grant from the Research Grants Council of Hong Kong (project code: HKU 704209P; HKU 704710P).
\end{acknowledgements}

\bibliographystyle{aa}
\bibliography{outflow.final}

\end{document}